\documentclass[12pt]{iopart}

\usepackage{cite}

\usepackage{graphicx}

\usepackage{dcolumn}

\begin{document}

\title[Energies for the Gaussian well]{Comment on: `A simple analytical expression for bound state energies for an
attractive Gaussian confining potential'}

\author{Francisco M Fern\'andez \ and Javier
Garcia}

\address{INIFTA (UNLP, CCT La Plata-CONICET), Divisi\'on Qu\'imica Te\'orica,
Blvd. 113 S/N,  Sucursal 4, Casilla de Correo 16, 1900 La Plata,
Argentina}\ead{fernande@quimica.unlp.edu.ar}

\maketitle

\begin{abstract}
We discuss a recently proposed analytical formula for the
eigenvalues of the Gaussian well and compare it with the
analytical expression provided by the variational method with the
simplest trial function. The latter yields considerably more
accurate results than the former for the energies and critical
parameters.
\end{abstract}

In a recent paper K\"{o}ksal\cite{K12} proposed a simple analytical
expression for the eigenvalues of the attractive Gaussian potential
\begin{equation}
V(r)=-\gamma e^{-\lambda r^{2}}  \label{eq:VGW}
\end{equation}
where $\gamma >0$ is the well depth and $\lambda >0$ determines its width.
From the expansion of the potential-energy function about its minimum $%
V(r)=-\gamma (1-\lambda r^{2}+\frac{\lambda ^{2}}{2}r^{4}-\ldots )$ and
perturbation theory the author derived an expansion for the energy of the
form
\begin{equation}
E_{nl}=E_{nl}^{HO}+\Delta E_{nl}^{(1)}+\Delta E_{nl}^{(2)}+\ldots
\label{eq:PT_series}
\end{equation}
where $n=0,1,\ldots $ and $l=0,1,\ldots $ are the radial and
angular-momentum quantum numbers, respectively. The first term $E_{nl}^{HO}$
is the sum of the minimum potential energy $-\gamma $ plus the harmonic
oscillation about this minimum.

It is well known that the perturbation series (\ref{eq:PT_series}) is
suitable for sufficiently deep wells (sufficiently great $\gamma $)\cite{F00}%
. In order to obtain a better expression K\"{o}ksal rewrote that
perturbation expansion in terms of an exponential function. Since
the author did not describe the general strategy clearly we
conjecture that the main idea is embodied in the following
expression
\begin{equation}
E_{nl}^{K}=\frac{1}{2}\left( E_{nl}^{HO}+\gamma \right) -\gamma e^{-\left(
E_{nl}^{HO}+\gamma \right) /(2\gamma )}  \label{eq:E_anal}
\end{equation}
For large $\gamma $ we expand the exponential function and obtain
the first term of the perturbation series (\ref{eq:PT_series})
exactly and the approximation $-\frac{1}{8\gamma }\left(
E_{nl}^{HO}+\gamma \right) ^{2}$ to the second one. K\"{o}ksal did
not discuss the agreement between the analytical formula
(\ref{eq:E_anal}) and the perturbation series (\ref
{eq:PT_series}). Consequently, without further justification this
expression can be considered to be an empirical formula and its
validity determined solely by the accuracy of the results. It is
worth noting that we can write equation~(\ref{eq:E_anal}) without
recourse to perturbation theory because we only need the term of
order zero.

K\"{o}ksal carried out some calculations for the particular model parameters
$\lambda =1/a_{B}^{2}$ and $\gamma =400\,Ryd$, where $a_{B}$ is the Bohr
radius and $Ryd$ the Rydberg energy. The approximate formula (\ref{eq:E_anal}%
) appears to approach the numerical eigenvalues reasonably well for some
values of the quantum numbers. However, we do not know the actual accuracy
of the empirical formula (\ref{eq:E_anal}) because the author did not report
results for other well depths. What we already know is that the accuracy of
the empirical formula decreases with $l$ and most remarkably with $n$\cite
{K12}.

The purpose of this comment is to test the accuracy of the empirical formula
(\ref{eq:E_anal}) more extensively and compare it with a simple analytical
expression obtained by means of the variational method.

The Schr\"{o}dinger equation is
\begin{eqnarray}
H\psi &=&E(\gamma ,\lambda )\psi  \nonumber \\
H &=&-\frac{\hbar ^{2}}{2m}\nabla ^{2}+V(r)  \label{eq:Schro}
\end{eqnarray}
where $m$ is the mass of the particle which K\"{o}ksal chose to be
the electron. It is always convenient to work with a dimensionless
eigenvalue equation that we easily derive in terms of the
dimensionless coordinates $\mathbf{r}^{\prime }=\mathbf{r}/L$,
where $L$ is an appropriate length unit. The Schr\"{o}dinger
equation thus becomes $H^{\prime }\psi ^{\prime }=E^{\prime }\psi
^{\prime }$, where $H^{\prime }=mL^{2}H/\hbar ^{2}$ and $E^{\prime
}=mL^{2}E/\hbar ^{2}$. If, for example, we choose $L=\lambda
^{-1/2}$ then we obtain
\begin{eqnarray}
H^{\prime } &=&-\frac{1}{2}\nabla ^{\prime 2}-\xi e^{-r^{\prime 2}}
\nonumber \\
\xi &=&\frac{m\gamma }{\lambda \hbar ^{2}}  \label{eq:Schro_dim}
\end{eqnarray}
where $\nabla ^{\prime 2}=L^{2}\nabla ^{2}$. Note that $E^{\prime }$ depends
only on the parameter $\xi $ since $E^{\prime }(\xi )=E(\xi ,1)=\xi E(\gamma
,\lambda )/\gamma $.

The dimensionless version of the empirical formula (\ref{eq:E_anal}) is
\begin{equation}
E_{nl}^{\prime \, K}(\xi )=\frac{1}{2}\left(
2n+l+\frac{3}{2}\right) \sqrt{2\xi }-\xi e^{-\frac{1}{2}\left(
2n+l+\frac{3}{2}\right) \sqrt{\frac{2}{\xi }}}
\label{eq:E_anal_dim}
\end{equation}
so that the discussion of its accuracy is greatly facilitated by
the fact that we need to vary just one model parameter. Note that
when $\lambda =1/a_{B}^{2}$ then $\xi =\frac{\gamma
}{2}\frac{2ma_{B}^{2}}{\hbar ^{2}}$ is half the well depth in
Rydberg units $Ryd=\frac{\hbar ^{2}}{2ma_{B}^{2}}$.
Therefore, the particular values of the model parameters $\gamma $ and $%
\lambda $ chosen by K\"{o}ksal correspond to $\xi =200$.

In a recent pedagogical article Fern\'{a}ndez\cite{F11a} discussed the
application of the variational method to the one-dimensional Gaussian well
(see also\cite{F12}). We can apply the same approach to the Gaussian well in
three dimensions. Following those papers we choose the simple trial function
\begin{equation}
\varphi (r)=Nr^{l+1}e^{-ar^{2}}  \label{eq:Phi_var}
\end{equation}
where $N$ is a normalization factor and $a>0$ is a variational parameter (we
drop the primes on the dimensionless variables from now on). The optimal
value of $a$ is given by a root of $d\left\langle H_{r}\right\rangle /da=0$,
where $H_{r}$ is the radial Hamiltonian
\begin{equation}
H_{r}=-\frac{1}{2}\frac{d^{2}}{dr^{2}}+\frac{l(l+1)}{2r^{2}}-\xi e^{-r^{2}}
\label{eq:H_r}
\end{equation}
We thus obtain

\begin{eqnarray}
\xi &=&\frac{\left( 2a+1\right) ^{\frac{2l+5}{2}}}{2^{l-\frac{1}{2}}4a^{%
\frac{2l+1}{2}}}  \nonumber \\
E_{0l}^{\prime \,var} &=&\frac{a\left( 2l+1-4a\right) }{2}
\label{eq:E_var}
\end{eqnarray}
We can proceed in two alternative ways: either, given $\xi $ we solve the
first equation numerically for $a$ and then obtain the energy or we obtain
both $\xi $ and the energy analytically for a set of values of $a$ (a
parametric equation for the energy).

\begin{figure}[h]
\begin{center}
\includegraphics[width=9cm]{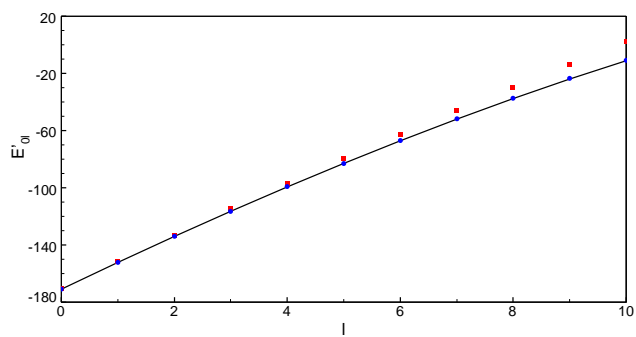}
\end{center}
\caption{Energy eigenvalues $E^{\prime\,K}_{0l}$ (red squares), $%
E^{\prime\,var}_{0l}$ (blue circles) and $E^{\prime\,exact}_{0l}$ (solid
line) for $\xi=200$}
\label{Fig:E_0L}
\end{figure}

Fig.~\ref{Fig:E_0L} shows the eigenvalues $E_{0l}^{\prime }$ for $\xi =200$
and several values of $l$ calculated by means of equations (\ref
{eq:E_anal_dim}) and (\ref{eq:E_var}). The highly accurate eigenvalues
provided by the Riccati-Pad\'{e} method (RPM)\cite{FMT89a} can be considered
to be exact for present purposes. As discussed above $\xi =200$ corresponds
to the potential parameters chosen by K\"{o}ksal. We appreciate that $%
E_{0l}^{\prime \,K}$ deviates from the exact result as $l$ increases. On the
other hand, the variational energy $E_{0l}^{\prime \,\,var}$ deviates so
less noticeably that it appears to agree exactly with the exact energy in
the scale of the figure.

\begin{figure}[h]
\begin{center}
\includegraphics[width=9cm]{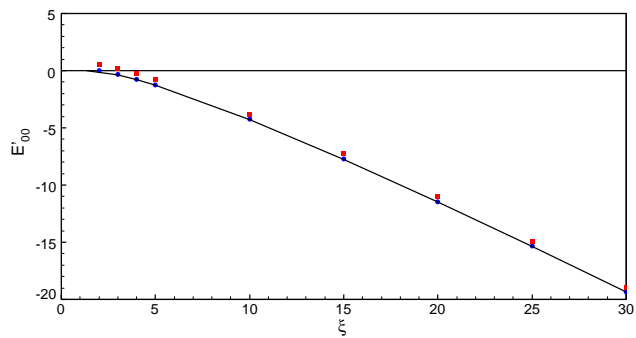}
\end{center}
\caption{Ground-state energy $E^{\prime\,K}_{00}$ (red squares), $%
E^{\prime\,var}_{00}$ (blue circles) and $E^{\prime\,exact}_{00}$ (solid
line) for a range of $\xi$ values}
\label{Fig:E_00}
\end{figure}

Fig.~\ref{Fig:E_00} compares the approximate ground-state energies
$E_{00}^{\prime \,K}$, $E_{00}^{\prime \,\,var}$ and the exact RPM
ones for a range of values of $\xi $. We appreciate that the
variational energy is closer to the exact one for all values of
$\xi $. However, the empirical expression (\ref {eq:E_anal_dim})
appears to yield reasonable results for the ground-state energy
for all those values of the well depth. Note that the largest
potential parameter $\xi =30$ in Fig~\ref{Fig:E_00} is
considerably smaller than the one chosen by K\"{o}ksal. It is well
known that the deepest the well the more accurate the results of
perturbation theory\cite{F00}. For this reason the values of the
potential parameters in Fig.~\ref{Fig:E_00} pose a good test for
any formula based on perturbation theory.

\begin{figure}[h]
\begin{center}
\includegraphics[width=9cm]{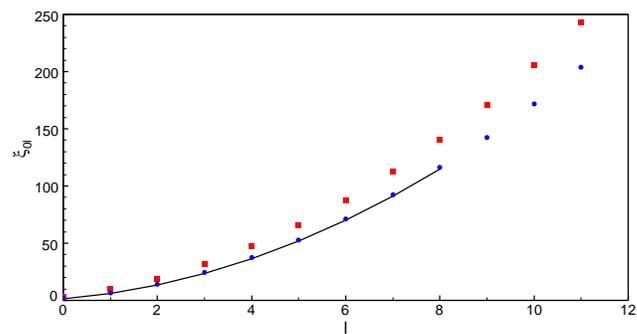}
\end{center}
\caption{Critical parameters $\xi^K_{0l}$ (red squares), $\xi^{var}_{0l}$
(blue circles) and $\xi^{exact}_{0l}$ (solid line)}
\label{Fig:xi_0l}
\end{figure}

It is also well known that the Gaussian well supports a finite number of
bound states and that there are critical values of the potential parameter $%
\xi $ for which bound states are exactly at the threshold of the continuum
spectrum $E=0$. In other words, there exists a bound-state eigenvalue $%
E_{nl}^{\prime }$ provided that $\xi >\xi _{nl}$, where $E_{nl}^{\prime
}(\xi _{nl})=0$. K\"{o}ksal did not discuss this important problem by means
of the empirical formula (\ref{eq:E_anal}) altough it is obvious that we can
obtain estimates $\xi _{nl}^{K}$ from the roots of $E_{nl}^{\prime \,K}(\xi
)=0$. We cannot solve this equation exactly but the numerical calculation is
simple enough. On the other hand, from the variational energy (\ref{eq:E_var}%
) we obtain $a_{0l}=(2l+1)/4$ and the extremely simple analytical formula

\begin{equation}
\xi _{0l}^{var}=\frac{\left( 2l+3\right) ^{\frac{2l+5}{2}}}{8\left(
2l+1\right) ^{\frac{2l+1}{2}}}  \label{eq:crit_par}
\end{equation}
Fig.~\ref{Fig:xi_0l} shows $\xi _{0l}^{K}$, $\xi _{0l}^{var}$ and the
accurate numerical results obtained by Liverts and Barnea\cite{LB11}. It is
clear that while $\xi _{0l}^{K}$ merely follows the trend $\xi _{0l}^{var}$
is almost indistinguishable from the exact results in the scale of the
figure. More precisely, the accuracy of $\xi _{0l}^{K}$ decreases noticeably
with $l$ while $\xi _{0l}^{var}$ remains remarkably accurate for all $l$
values.

\begin{figure}[h]
\begin{center}
\includegraphics[width=9cm]{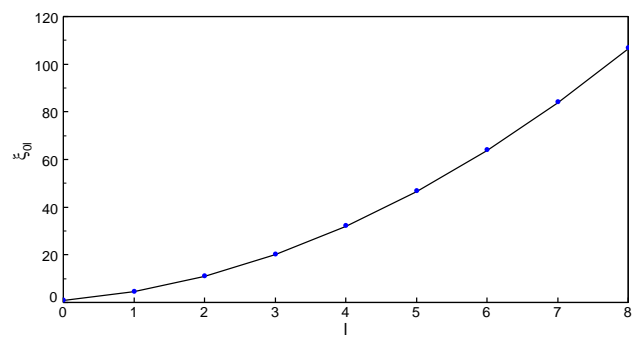}
\end{center}
\caption{Critical parameters $\xi^{var}_{0l}$ (blue circles) and $%
\xi^{exact}_{0l}$ (solid line)}
\label{Fig:xi_0l_Y}
\end{figure}

Since the author did not give a clear justification for the empirical
formula (\ref{eq:E_anal}) nor a sound procedure that may be applied to other
problems we assume that the sole purpose of the paper was to obtain an
empirical formula for the eigenvalues of the Gaussian well. This assumption
is supported by the fact that K\"{o}ksal did not attempt to derive a similar
expression for the eigenvalues of the Yukawa potential already treated by
the same perturbation method in an earlier paper\cite{GKB06}. On the other
hand, the variational method discussed above is not restricted to the
Gaussian well and can be easily applied to the dimensionless Schr\"{o}dinger
equation for the Yukawa potential
\begin{equation}
V(r)=-\frac{\xi }{r}e^{-r}  \label{eq:V(r)_Y}
\end{equation}
By means of the trial function
\begin{equation}
\varphi (r)=Nr^{l+1}e^{-ar}
\end{equation}
we obtain the variational parametric formula for the energy

\begin{eqnarray}
\xi &=&\frac{\left( l+1\right) \left( 2a+1\right) ^{2l+3}}{2^{2\left(
l+1\right) }a^{2l+1}\left( 2a+2l+3\right) }  \nonumber \\
E_{0l}^{\prime \,var} &=&\frac{a^{2}\left( 2l+1-2a\right) }{2\left(
2a+2l+3\right) }  \label{eq:e_var_Y}
\end{eqnarray}
and the critical parameters are given by the simple analytical expression

\begin{equation}
\xi _{0l}^{var}=\frac{2^{2l}\left( l+1\right) ^{2l+3}}{\left( 2l+1\right)
^{2l+1}}  \label{eq:crit_par_Y}
\end{equation}
Fig.~\ref{Fig:xi_0l_Y} shows the remarkable agreement between this formula
and the accurate numerical results of Liverts and Barnea\cite{LB11}.

Finally, we summarize the main conclusions of this comment:

First, K\"{o}ksal's empirical formula is far less accurate than the
analytical expression provided by the simplest variational function. It is
true that K\"{o}ksal's formula applies to states with $n>0$ while the
variational method does not yield simple analytical expressions for such
states (the Rayleigh-Ritz method suitable for them should be treated
numerically). However, it is also true that K\"{o}ksal's empirical formula
becomes considerably less accurate as $n$ increases\cite{K12} and here we
have just compared the results for the most favourable case $n=0$.

Second, the variational method applies to other problems as we
have just illustrated by means of the Yukawa potential. For
unknown reasons K\"{o}ksal did not attempt to apply the same
approach to other models for which perturbation corrections are
already available\cite{F00} as it is the case of the Yukawa
potential\cite{GKB06}.

\end{document}